\documentclass[preprint]{aastex}

\newcommand{\angst} {\mbox{$\:${\rm \AA}}}
\newcommand{\micr}  {\mbox{$\,\mu{\rm m}$}}
\newcommand{\cm}  {\mbox{$\:{\rm cm}$}}
\newcommand{\mm}  {\mbox{$\:{\rm mm}$}}
\newcommand{\m}   {\mbox{$\:{\rm m}$}}

\begin{document}

\title{A Large Area CCD Camera for the Schmidt Telescope at the Venezuelan
National Astronomical Observatory}

\author{C.~Baltay\altaffilmark{1}, J.~A.~Snyder\altaffilmark{1,2},
P.~Andrews, W.~Emmet, B.~Schaefer\altaffilmark{3}, J.~Sinnott\altaffilmark{4}}
\affil{Physics Department, Yale University, New Haven, CT, 06520--8121, USA}
\email{charles.baltay@yale.edu, jeffrey.snyder@yale.edu,
peter.andrews@yale.edu, william.emmet@yale.edu, schaefer@astro.as.utexas.edu,
jps39@cornell.edu}

\author{C.~Bailyn, P.~Coppi\altaffilmark{5}, A.~Oemler\altaffilmark{6},
C.~N.~Sabbey\altaffilmark{7}, S.~Sofia, W.~van~Altena, A.~K.~Vivas}
\affil{Astronomy Department, Yale University, New Haven, CT, 06520--8101, USA}
\email{bailyn@astro.yale.edu, coppi@astro.yale.edu, oemler@ociw.edu,
cns@boglefunds.com, sofia@astro.yale.edu, vanalten@astro.yale.edu,
vivas@astro.yale.edu}

\author{C.~Abad, C.~Brice\~{n}o, G.~Bruzual, G.~Magris, J.~Stock,
F.~Della~Prugna, Ge.~S\'{a}nchez, Gu.~S\'{a}nchez, H.~Schenner}
\affil{Centro de Investigaciones de Astronom\'{\i}a, M\'{e}rida, Venezuela}
\email{abad@cida.ve, briceno@cida.ve, bruzual@cida.ve, magris@cida.ve,
stock@cida.ve, dellap@cida.ve, gerardo@cida.ve, sanchez@cida.ve}

\author{B.~Adams, M.~Gebhard, R.~K.~Honeycutt, J.~Musser}
\affil{Indiana University, Bloomington, IN, 47405--7105, USA}
\email{adams@astro.indiana.edu, gebhard@astro.indiana.edu,
honey@astro.indiana.edu, musser@astro.indiana.edu}

\author{F.~Harris\altaffilmark{8}}
\affil{Universities Space Research Association, USNO, Flagstaff, AZ, 86002--1149, USA}
\email{fhh@nofs.navy.mil}

\and

\author{J.~Geary}
\affil{Harvard-Smithsonian Center for Astrophysics, 60 Garden Street,
Cambridge, MA 02138, USA}
\email{jgeary@cfa.harvard.edu}

\altaffiltext{1}{Also Astronomy Department, Yale University, New Haven, CT,
06520--8101, USA}
\altaffiltext{2}{Send correspondence to jeffrey.snyder@yale.edu}
\altaffiltext{3}{Present location University of Texas, Austin, TX, 78712, USA}
\altaffiltext{4}{Present location Cornell University, Ithica, NY, 14853-1501,
USA}
\altaffiltext{5}{Also Physics Department, Yale University, New Haven, CT,
06520--8121, USA}
\altaffiltext{6}{Also Carnegie Observatories, Pasadena, CA, 91101, USA}
\altaffiltext{7}{Present location: Bogle Investment Management, Wellesley, MA,
02481, USA}
\altaffiltext{8}{Present location: U.~S.~Naval Observatory, Flagstaff, AZ,
86002--1149, USA}

\begin{abstract}

We have designed, constructed and put into operation a large area CCD
camera that covers a large fraction of the image plane of the 1 meter
Schmidt telescope at Llano del Hato in Venezuela.  The camera consists
of 16 CCD devices arranged in a $4 \times 4$ mosaic covering $2.3\degr
\times 3.5\degr$ of sky.  The CCDs are $2048 \times 2048$ LORAL
devices with $15\micr$ pixels.  The camera is optimized for drift scan
photometry and objective prism spectroscopy.  The design considerations,
construction features and performance parameters are described in the
following article.

\end{abstract}
\keywords{instrumentation: detectors, surveys, cosmology}

\section{Introduction}
\label{sec:intro}

Schmidt telescopes are the instrument of choice for surveys of large
areas of the sky because of their large field, typically of the
order of $4\degr \times 4\degr$.  However, these telescopes have large,
curved image planes and are difficult to instrument with silicon detectors.
Until now, these telescopes have been used with photographic plates
and no Schmidt telescope had its image plane fully instrumented with
silicon CCD detectors.  The Near--Earth Asteroid Tracking (NEAT) project
has recently instrumented the Palomar 48'' Oschin Schmidt telescope with
three $4080 \times 4080$ CCDs \citep{Pravdo2002}.  This camera covers
a total of $\sim 3.75$ square degrees of the $\sim 36$ square degree
field of view.  A project is underway to fully instrument this telescope.

\begin{figure}
  \vspace*{7in}
  \includegraphics{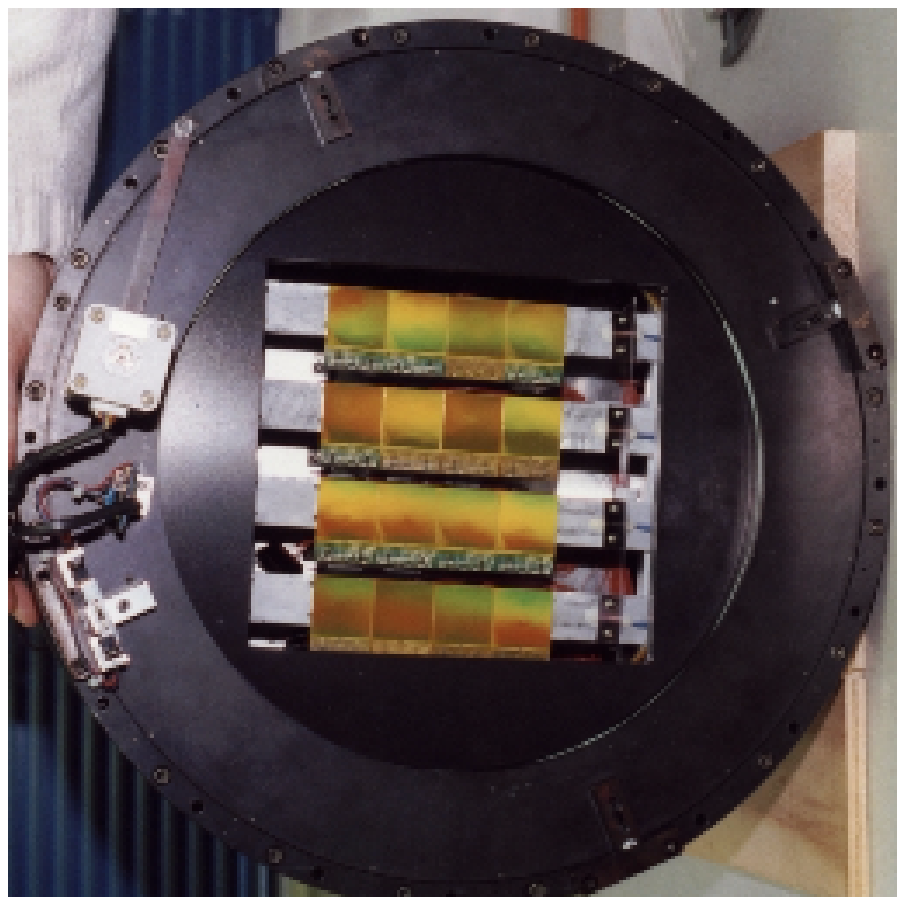}
  \caption{A picture of the QUEST camera.  The drift direction is from the
top towards the bottom.}
  \label{f:camera_pic}
\end{figure}

\begin{figure}
  \vspace*{5in}
  \includegraphics{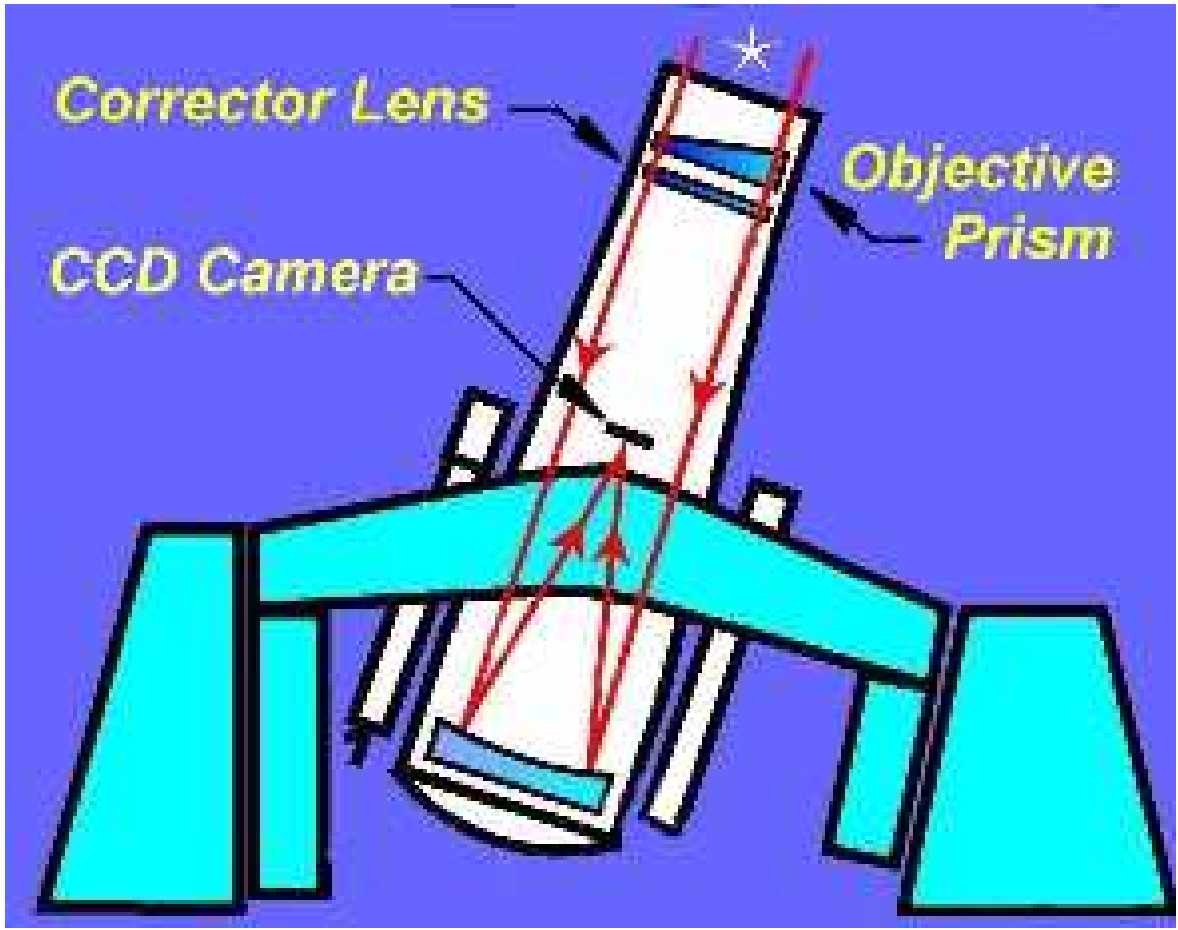}
  \caption{The Venezuelan Schmidt telescope showing the QUEST camera at the
prime focus.  Note that the prism angle is exaggerated and the bending of
the input rays is neglected.}
  \label{f:telescope}
\end{figure}

We, the QUEST collaboration,\footnote{QUEST is short for the Quasar
Equatorial Survey Team, and is a collaboration between groups from
Yale University, Indiana University, Centro de Investigaciones de
Astronom\'{\i}a (CIDA), and Universidad de Los Andes (M\'erida,
Venezuela).} have designed, constructed and put into operation, a
large area CCD camera that covers a large fraction of the image plane
of the 1 meter Schmidt telescope at the Venezuelan National
Astronomical Observatory located at Llano del Hato and operated by
CIDA.\footnote{Research reported herein is based on data obtained with
the 1m Schmidt telescope at the Venezuelan National Astronomical
Observatory, M\'erida, Venezuela, operated by the Centro de
Investigaciones de Astronom\'{\i}a (CIDA) and funded by the Ministerio
de Ciencia y Technolog\'{\i}a and the Fondo Nacional de Ciencia y
Technolog\'{\i}a of Venezuela.}  This is one of the five largest
Schmidt telescopes in the world; its properties are summarized in
table~\ref{t:properties}.  A picture of the camera is shown in
figure~\ref{f:camera_pic}.  The camera is located at the prime focus
inside the telescope tube as shown in figure~\ref{f:telescope}.

The scientific motivation for  building this camera was to carry out
a large area survey of a band of the sky centered on the celestial
equator and about $\pm6\degr$ wide in declination.  The initial plan
was to divide the observations between objective prism spectroscopy,
imaging with four essentially simultaneous color filters, and repeated
scans of a 250 square degree area of the sky for variability studies.
Some of the scientific results expected from such a survey were a quasar
survey based on quasar selection with three different techniques:
selection by the presence of broad emission lines using the objective prism,
color selection with UBV and BVR colors, and variability selection.  The
variability survey was included to search for type Ia supernovae, gamma
ray bursters, new solar system objects (like asteroids and Kuyper belt
objects), and RR Lyrae stars.

\begin{figure}
  \vspace*{4.5in}
  \includegraphics{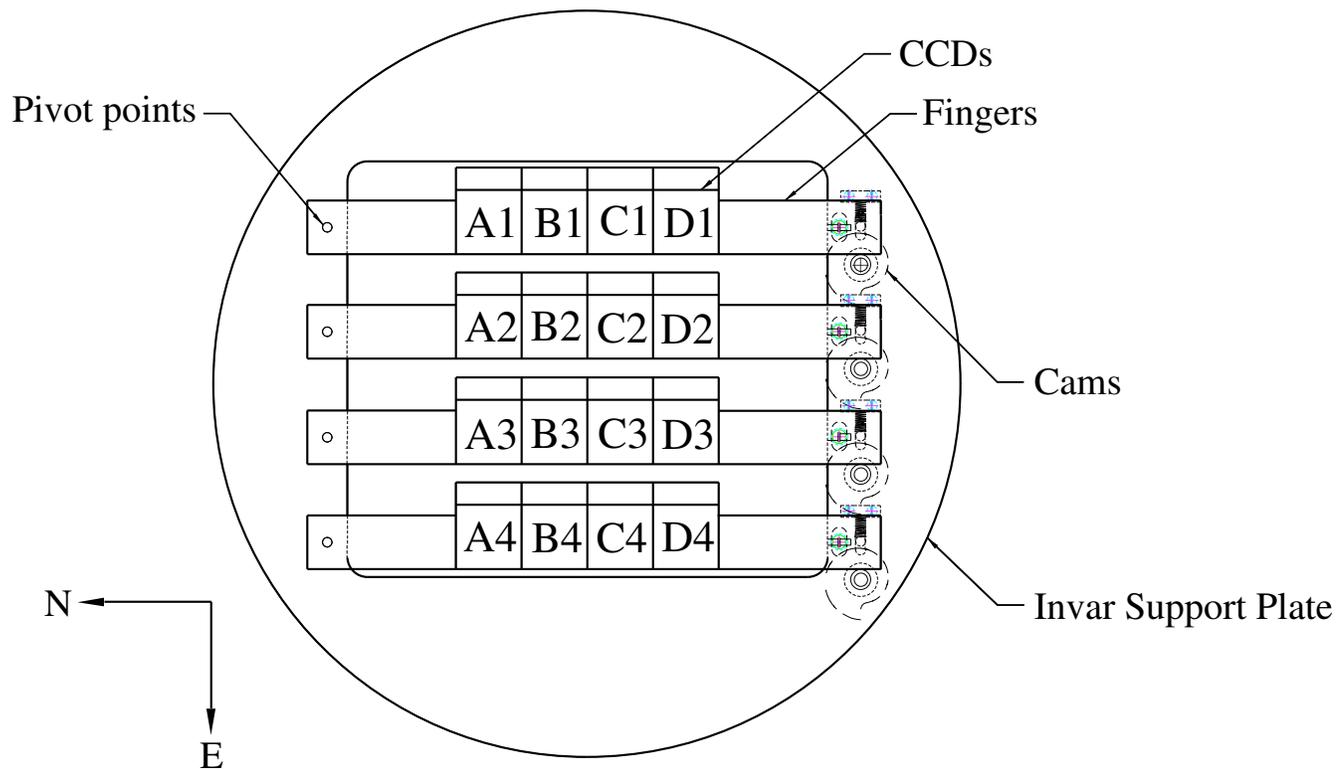}
  \caption{Layout of the CCDs on the image plane.  Also shown are the Invar
fingers supporting the CCDs, their pivot points and the finger rotating cams.}
  \label{f:camera_front}
\end{figure}

The large area CCD camera consists of 16 CCD devices arranged in a $4
\times 4$ mosaic (see fig.~\ref{f:camera_front}) covering $2.3\degr
\times 3.5\degr$ of sky.  The individual CCDs are $2048 \times 2048$
LORAL devices with $15\micr$ pixels.  There are gaps between the CCDs
in the East--West direction so that the effective area covered is 5.4
square degrees.  The properties of the camera are summarized in
table~\ref{t:camera_prop}.  The camera has been designed to operate in
drift scan mode,\footnote{We thank Steve Shectman for many interesting
discussions concerning the technique of drift scanning.}
 which is also referred to as time--delay integration
(TDI) mode \citep{McGraw1986}.  In this mode the telescope is locked
into a fixed position at a given declination angle.  The CCD array is
oriented with the columns of pixels in the clocking direction lined up
precisely in the East--West direction, and the CCDs are clocked
synchronously with the motion of the star images across each CCD.
Each star image thus crosses four CCDs, one in each row of CCDs.  Each
row of CCDs can have a filter of a different color in front of it so
the camera can collect images in each of four colors essentially
simultaneously.  The camera thus has a 100\% duty cycle (i.e., data is
collected continuously) and no time is lost due to readout time or
telescope slewing time.  Since the telescope is locked into fixed
position and is not tracking, the system is very stable and produces
more accurate photometric measurements.  Photometric
precision is further enhanced because each point in the sky is imaged
by averaging over an entire column of pixels and thus pixel--to--pixel
variations in sensitivity are minimized.

During a clear night, a 2.3\degr\ wide by 120\degr\ long strip, or
approximately 250 square degrees, can be covered in each of four
colors.  The telescope is equipped with an objective prism covering
the full aperture of the telescope and can be installed or removed on
a daily basis.  Thus, one can collect objective prism spectra over 250
square degrees of sky on a clear night.  The camera was built at Yale
University and Indiana University.  It was installed at the prime
focus of the telescope in 1997 and has been taking data routinely
since November 1998.
	
In the following sections, we will describe the principles of
operation of the camera (\S\ref{sec:operation}), give a detailed
description of the camera (\S\ref{sec:description}), the detector
control system and electronics (\S\ref{sec:electronics}), the data
analysis software (\S\ref{sec:software}), and the performance of
the apparatus to date (\S\ref{sec:performance}).

\section{Principles of Operation of the Camera}
\label{sec:operation}

\begin{figure}
  \vspace*{3.5in}
  \includegraphics{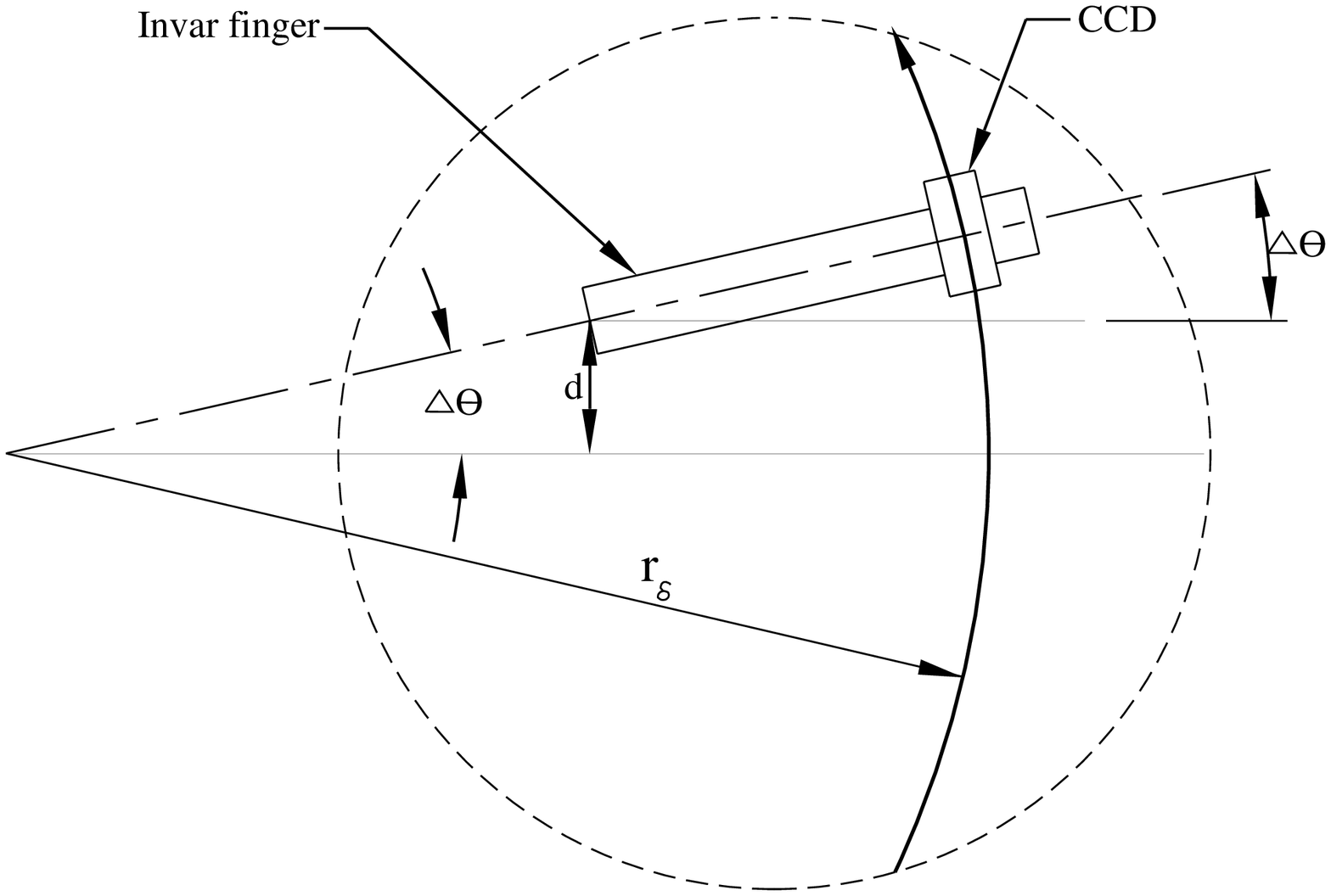}
  \caption{Schematic diagram of the CCD rotation to keep each
CCD lined up along the line of motion of the star images.}
  \label{f:CCD_rot}
\end{figure}

If the camera were used to drift scan along the equator, the images of
stars would follow straight lines and move at the same rate across the
image plane.  However, at declinations other than the equator, the
stars will follow arcs of circles and stars at different North-South
positions will move at different rates.  In drift scanning, the
sagittas due to the first effect will smear the images in the
North-South direction, and due to the second effect some of the stars
will not be exactly synchronous with the CCD clocking rates and thus
will be smeared in the East-West direction.  In order to keep these
effects at an acceptable level, i.e. to keep the smearing of the point
spread function below about one arc second in any direction, we rotate
each CCD by an amount dependent on the declination being scanned in
such a way that the clocking direction of each CCD is tangential to
the arcs that the stars are moving in at that location in the array.
This is accomplished by mounting each of the four CCDs in a
North-South row on an Invar\footnote{Invar is a stainless steel alloy
that has a relatively low coefficient of thermal expansion.} bar,
which we call a ``finger.''  Each of the four fingers can be rotated
by a different amount by cams which are driven by external, computer
controlled stepper motors.  An exaggerated sketch of this scheme
is shown in figure~\ref{f:CCD_rot}.  For convenience, we label the
fingers 1, 2, 3, and 4, and the columns of CCDs A, B, C, and D, as
shown in figure~\ref{f:camera_front}.  This figure also shows the
pivot points and the cams used to rotate the fingers.  In addition,
each column of CCDs is scanned along a slightly different
declination, and therefore, the parallel clocks reading out the CCDs
are synchronized at slightly different rates.

The radius $r_{\delta}$ of the star tracks (i.e. the arcs of
circles along which the image of a star moves in drift scanning) on
the image plane of a telescope with focal length {\em f} at a
declination $\delta$ is to a good approximation given by:
$$r_{\delta}=\frac{f}{\tan \delta}.$$
The parallel clock rate $\nu_{\delta}$ for reading out a CCD in such a
way that the motion of the charge is synchronous with the motion of
the star image across the CCD at a declination $\delta$ is given by:
$$\nu_{\delta} = \frac{\Omega f}{a}\cos \delta.$$
where $\Omega=72.7 \mu{\rm radians/sec}$ is the rotation rate of the Earth,
$f$ is the focal length of the telescope, and $a$ is the pixel
size on the CCD.  For the Venezuelan Schmidt telescope and our pixel size,
this gives:
$$\nu_{\delta} = 14.7 \cos \delta \qquad {\rm lines/second}.$$
	
In drift scanning along the equator, the readout parallel clocks are
thus synchronized at approximately 14.7 lines/second.  At this rate, a
star image takes 140 seconds to cross a CCD.  This gives an
integration or exposure time of 140 seconds.  At higher declinations,
the clocking rate is somewhat slower giving a slightly longer exposure
time.  In drift scan mode, this exposure time is governed by the
rotation of the earth and can not be changed.  However, since each
star crosses four CCDs, these can be added for an effective exposure
time of 560 seconds.  In cases where even longer exposure times are
desirable, repeated scans of the same area of sky can be performed and
co--added.
	
The angle by which the CCD support fingers (see figure~\ref{f:CCD_rot}) have
to be rotated to keep the clocking direction of the CCDs tangent to
the star tracks on the image plane at a declination $\delta$ is:
$$\Delta\theta = \frac{d}{f}\tan\delta $$
where $d$ is the distance of the pivot point of each finger from
the camera centerline ($d = \pm 7.5\cm$ for fingers 1 and 4 and
$\pm 2.5\cm$ for fingers 2 and 3).  Thus, for example, the top finger
(1) has to be rotated about $0.15\degr$ for $\delta=6\degr$, which is very
small but nevertheless quite important to keep the image sizes small.
As mentioned above, the four columns of CCDs scan along slightly
different declinations and thus have to be clocked at slightly
different rates.  For example, with the camera scan centered at
$\delta=6\degr$, the four columns have to be clocked at 14.638, 14.626,
14.613, and 14.598 lines/second, respectively. Again this variation is
very small but nevertheless quite important.  The detailed
implementation of this scheme is to clock all four columns at the same
rate but drop clock pulses at different rates in the four columns to
produce the average clock rates required.  The fast serial
readout clock is 50 kHz for all of the CCDs in the array.

\begin{figure}
  \vspace*{2.5in}
  \includegraphics{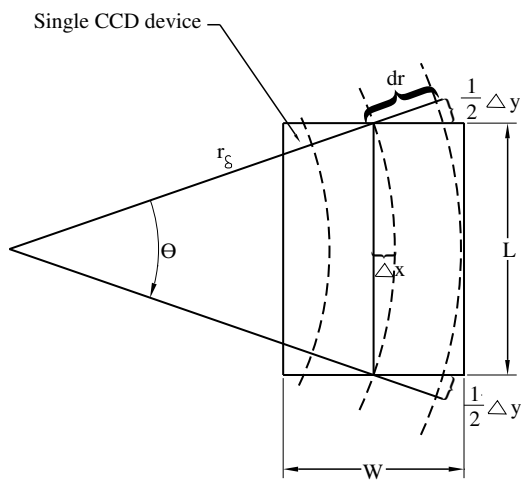}
  \caption{Sagitta ($\Delta x$) and path length difference ($\Delta y$)
on a single CCD.}
  \label{f:sagitta}
\end{figure}

The scheme described above for varying the rotation and clock rate
synchronization of the different CCDs in the array removes the
dominant effects that smear the images.  There are, however, 
residual effects due to the sagitta of the image motion and spread
in the rate of motion of
the images across the finite width of a single CCD, as illustrated in
figure~\ref{f:sagitta}.  For a CCD with length {\em l} (in the E-W
direction) and width {\em w} (in the N-S direction), the residual
smearing of the image size $\Delta x$ and $\Delta y$, in the E-W and
N-S directions respectively, scanning at a declination $\delta$, is
given by:
$$\Delta x = \frac{1}{8} \frac{l^{2}}{f}\tan \delta$$
$$\Delta y = \frac{1}{2} \frac{lw}{f}\tan \delta$$
where {\em f} is the focal length of the telescope.  This residual
smearing limits the range of declinations at which this camera can be
used in the drift scanning mode without intolerable degradation of the
image sizes.

The design has been optimized in such a way that the residual image
smearing is kept below 1 arcsec for declinations up to $\pm 6\degr$.
Given the typical seeing at the Llano del Hato site, we can drift scan
at declinations up to $\pm 12\degr$ with no appreciable degradation of
image quality.  This is sufficient for the equatorial survey for which
the camera has been designed.  Of course, the camera can be operated
in a conventional point and stare mode to cover regions of the sky
above these declinations.

Another complication of the design was due to the fact that the image
plane of a Schmidt telescope is not flat but has the shape of a convex
spherical surface.  To arrange the CCDs in such a shape would have
been cumbersome.  Instead, we designed, and had built, a 30\cm\ 
diameter field flattener lens that covered the entire image plane.
This lens produced a flat image plane and, in addition, corrected for
the pincushion distortion inherent in the telescope to a level where
the degradation of the image shapes were negligible.
	
The depth of field of the Venezuelan Schmidt telescope is quite shallow.
For this reason, the front surfaces of the 16 CCDs, including the
motion of the finger mounts, had to be kept in a plane to a tolerance
of less than $\pm 25$\micr.  It required great care in the precision
machining and the alignment procedures to achieve this precision.
	
During the commissioning period, after the camera had been installed in the
telescope, a great deal of effort was expended to align the plane of
the CCDs with the focal plane of the telescope.  Once this had been
achieved, however, it was quite stable and required no further
adjustment.  Typically, before each nights' data taking the focus of
the telescope, the rotational position of the fingers, and the
synchronized read out rate, which have been set by the control
computers for the appropriate declination, were checked by looking at
the shape and size of stellar images.

There is an objective prism designed for the Venezuelan Schmidt telescope
which covers the entire 1m aperture.  By aligning the dispersion
direction of the prism along the drift direction, we can obtain
spectra over the entire survey area.  The prism is made of UBK7 for
good ultraviolet transmission down to the atmospheric cutoff.  With a
prism angle of 3.3\degr, the dispersion is 650\angst/mm at 4350\angst.
In combination with the QUEST camera, this corresponds to a spectral
resolution of 10\angst\ per pixel at this wavelength which is more
than sufficient to detect quasars.

\section{Description of the QUEST Camera}
\label{sec:description}

The CCD camera is located at the prime focus of the Venezuelan Schmidt,
inside the telescope about 3\m\ from the 1.5\m\ primary mirror, with
the CCDs facing the mirror (see figure~\ref{f:telescope}).  The
outside dimensions of the camera have to be as small as possible to
fit mechanically at the prime focus and to obscure as small a part of
the incoming beam as possible.  The outside of the camera has to be
kept at ambient temperature so as not to induce any turbulence in the
air inside the telescope.

\subsection{The CCDs}

\begin{figure}
  \vspace*{4in}
  \includegraphics{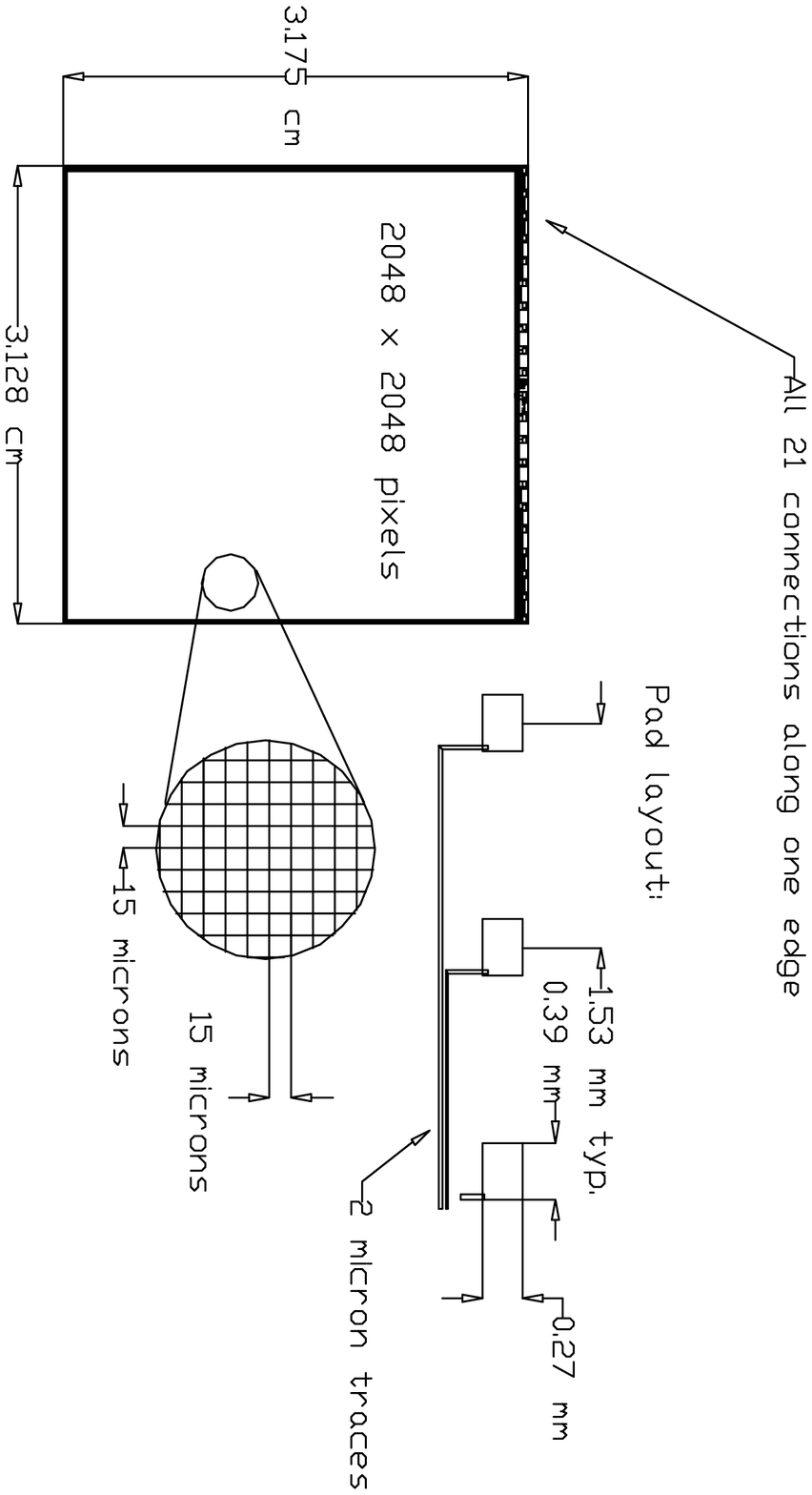}
  \caption{Sketch of the CCDs with $2048 \times 2048$
$15\micr \times 15\micr$ pixels.}
  \label{f:CCD_detail}
\end{figure}

The heart of the camera is a mosaic of 16 CCDs arranged as shown in
figure~\ref{f:camera_front}.  Each CCD consists of $2048 \times 2048$
pixels, each $15\micr \times 15\micr$, as shown in
figure~\ref{f:CCD_detail}.  The CCDs were designed by John Geary of
the Harvard--Smithsonian Center for Astrophysics (CfA) and were part
of a multi--lot fabrication run by LORAL (20 four inch diameter wafers
per lot with four CCDs per wafer).  These fabrication lots were
initially purchased by various U.S. and European observatories and we
were able to obtain a few individual unused devices from a number of
these observatories.\footnote{We are grateful to John Geary (CfA),
Edgar Smith (EST), Jerry Lupino (Hawaii), Kent Cook (Los Alamos) and
Fabio Bartoletto (Padua) for these devices.}  Most of the devices were
three--side buttable with 21 readout pads along a single edge; a few
of them were two--side buttable.  The two--side buttable CCDs could be
used on the outside columns (A and D) with no interference.  Each of
these devices had a number of bad pixels and some bad columns but
this does not significantly affect their performance.

\begin{figure}
  \vspace*{2in}
  \includegraphics{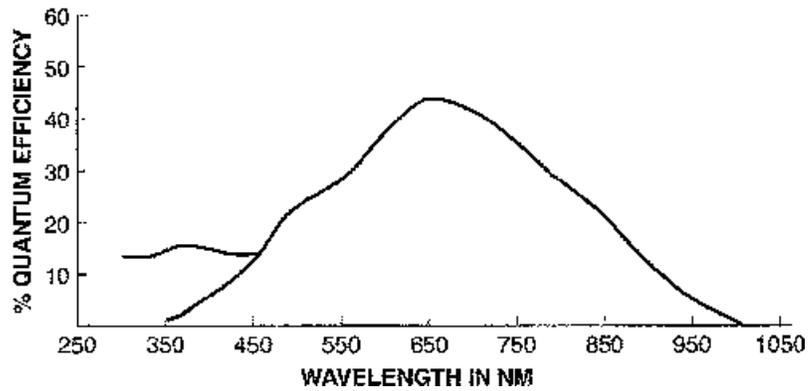}
  \caption{Typical quantum efficiency of front illuminated commercial
Loral CCDs.  The response below 4000\angst\ is due to a
wavelength shifting compound on the front surface.}
  \label{f:QE}
\end{figure}

These devices are unthinned (500\micr\ thick) and are used in
the front illluminated mode.  As such, the quantum efficiency drops
to zero below 4000\angst.  Since we need sensitivity in the
ultraviolet we applied a 2.5\micr\ thick layer of a wavelength
shifting compound to the front surface of each device.
This compound absorbs light in the UV and reemits
it in the visible.  We have not carried out detailed quantum efficiency
measurements but we believe that the QE is similar to the typical
response of front illuminated LORAL CCDs, as shown in figure~\ref{f:QE}
\citep[Adapted from][]{loral}, with
the modification at short wavelength due to the wavelength shifting
compound.  The properties of the CCDs are summarized in
table~\ref{t:CCD_prop}.

\begin{figure}
  \vspace*{3.4in}
  \includegraphics{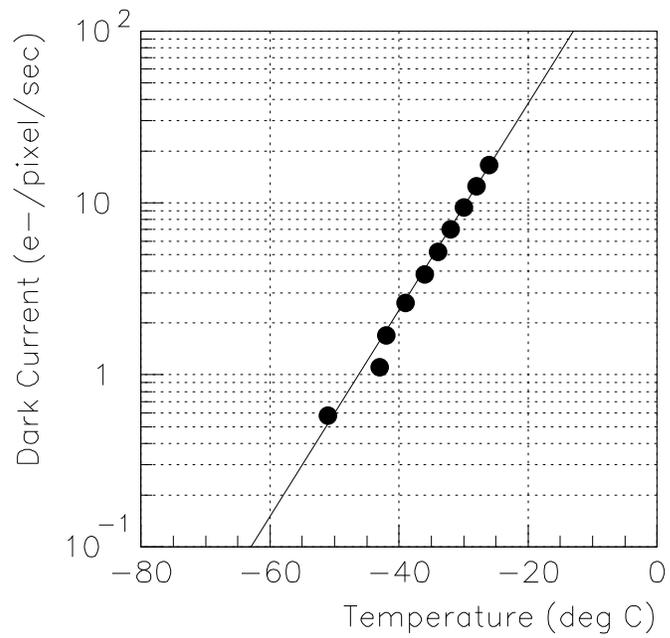}
  \caption{Typical dark current versus temperature plot for the CCDs.}
  \label{f:dark_current}
\end{figure}

The read noise of the CCDs, including the noise from the electronics
and pickup in the cabling, etc., is around 10 electrons, but with some
spread.  The best we have seen is 7 electrons noise, and some of the
devices in the camera are as poor as 20 electrons noise. The
measurement of the dark current versus temperature on a typical CCD is
shown in figure~\ref{f:dark_current}.  The dark current extrapolates
to $\sim$ 250 pA/cm$^{2}$ at 20\degr C, and falls a factor of 2 with
with each 5\degr C temperature drop.  The CCDs in the camera are
typically operated at $-70$\degr C.  The response of the CCDs is
linear to a good approximation up to the full well capacity, which
varies from 30,000 to over 60,000 electrons.

\subsection{The Field Flattener Lens}

\begin{figure}
  \vspace*{4in}
  \includegraphics{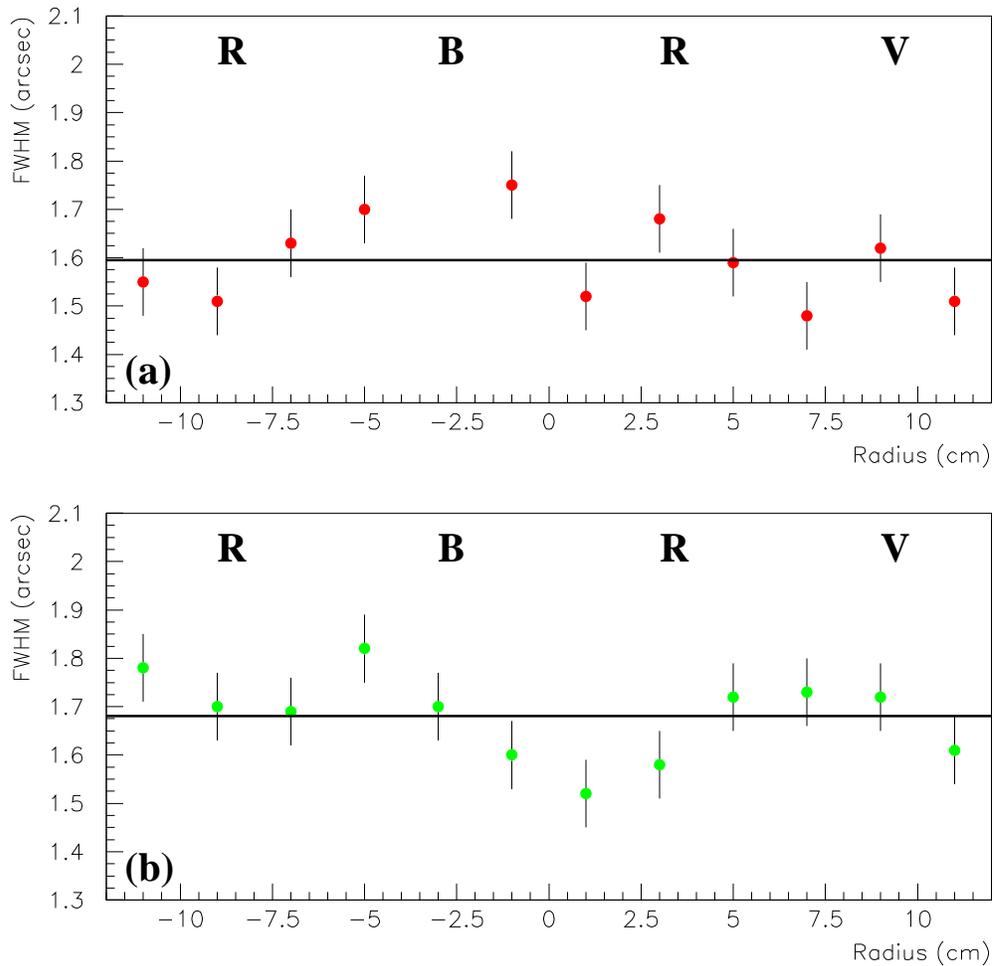}
  \caption{The average stellar full width at half maximum (FWHM) along 
the two diagonals of the full image plane for a typical exposure:
(a) across CCDs A1, B2, C3 and D4 and (b) across CCDs A4, B3, C2 and D1.
The fifth point in (a) is omitted since there were insufficient stars in
the data.}
  \label{f:FWHM}
\end{figure}

The Venezuelan Schmidt normally has a focal plane that has the shape of a
convex spherical surface.  To allow the CCD array to be in a flat
plane we have designed a field flattener lens to reimage the focal
plane.  Great care had to be exercised in this design to keep
pincushion and other distortions below a few microns in the image
plane to allow the camera to be used in a drift scan mode. The lens
was designed by the QUEST collaboration and was manufactured by Coastal
Optical Systems of Riviera Beach, Florida.  The lens has a 30\cm\
diameter and is also used as the vacuum window at the front of the
detector.  The lens is biconvex, with two spherical surfaces of radius
of curvature of $2595\mm$ and $-1725\mm$ respectively.  It is 25\mm\
thick in the center and 14\mm\ thick at the edges.  Careful finite
element stress analysis calculations have been carried out to show
that the lens is strong enough to serve as a vacuum window and the
deflection is small enough not to distort the optical properties of
the lens unduly.  The lens is made of fused silica, Corning 7940,
which has good transmission in the ultraviolet as well as over the
whole required wavelength range.  The lens does not have an
antireflection coating.  The lens is located about 2\cm\ in front of
the CCD image plane.  We have examined the image sizes and shapes at
different locations on the image plane.  No variation or degradation
as a function of distance from the center of the image plane has been
observed (see figure~\ref{f:FWHM}).  We therefore believe that this
lens has flattened the field to an acceptable level.

\subsection{The Camera Dewar}

\begin{figure}
  \vspace*{5.2in}
  \includegraphics{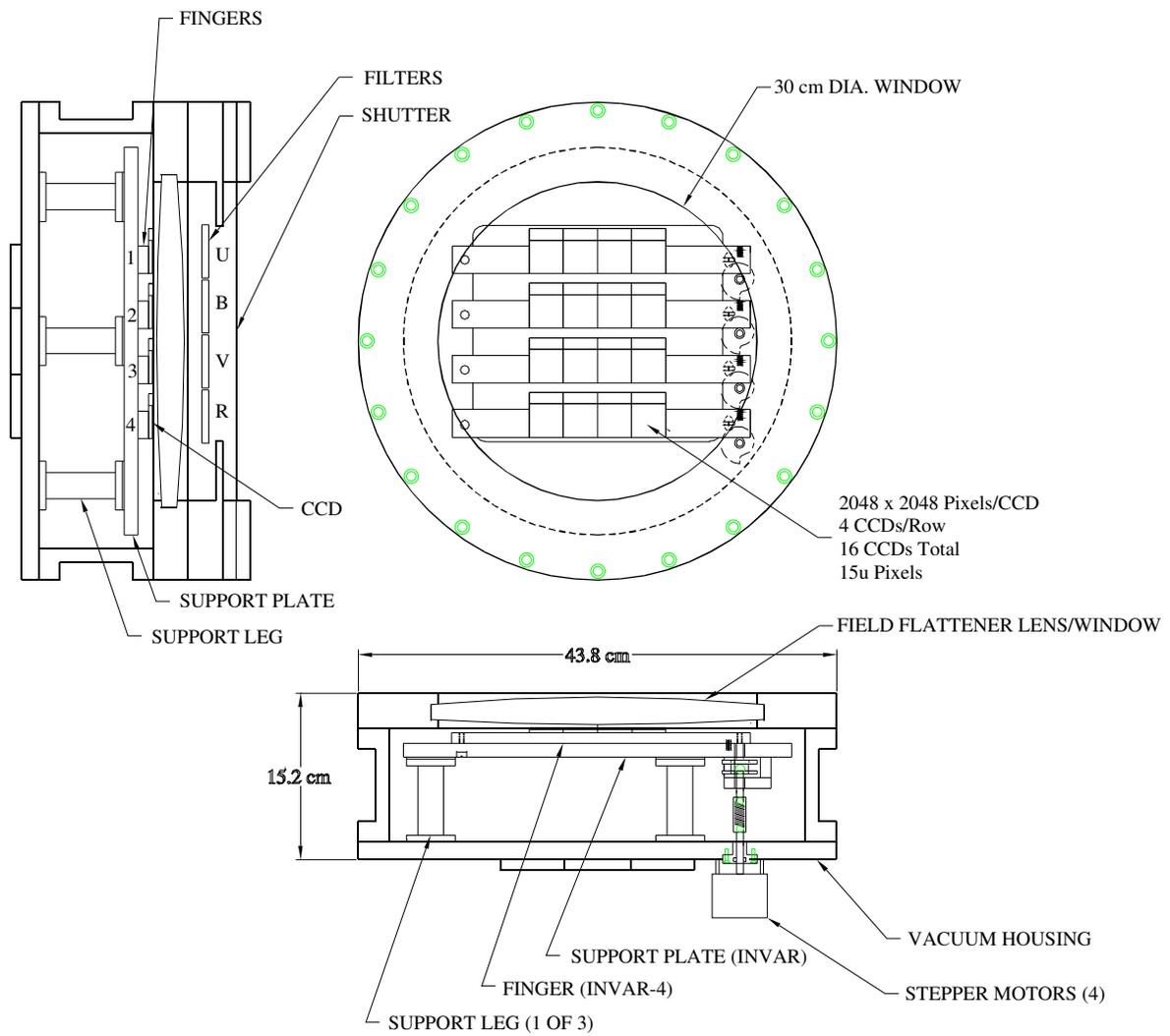}
  \caption{Frontal view and two cross--sectional views of the QUEST camera.}
  \label{f:camera_dewar}
\end{figure}

To be able to cool the CCDs to $-70$\degr C they have to be in a
vacuum enclosure.  The detector housing is thus a vacuum vessel
cylindrical in shape 16 inches in diameter, about 5 inches deep as
shown in figure~\ref{f:camera_dewar}.  The front face has a 12 inch
diameter circular vacuum window made of fused silica for good UV
transmission.  This window also serves as the field flattener lens
discussed above.  The back plate of the housing has the mounting
bracket on it to attach the detector to the telescope.  The back plate
also has penetrations for the vacuum pump port, the vacuum gauge,
refrigerant liquid in and out, and all of the electrical feedthroughs.
The housing is made of aluminum, polished on the inside to reduce
radiative heat loss, and anodized flat black on the outside to reduce
reflections and glare.

\subsection{Refrigeration}

The conventional method of cooling the CCDs in astronomical cameras is
to have a liquid nitrogen dewar inside the camera vacuum
enclosure.  In our case, this would have been a feasible solution but
would have made the detector quite bulky because of the large volume
of liquid needed to cool the large CCD array, and would present a
logistic problem since at this time there is no liquid nitrogen
available at the observatory.  We have found a much simpler and
maintenance free solution, namely a closed loop refrigerator using a
liquid freon coolant that can operate in the $-60$ to $-80$\degr C
temperature range.  Such a refrigerator was commercially available, a
model RC210C0 from FTS Systems in Stone Ridge, New York.  The
refrigeration unit is located near the North pier of the
Schmidt telescope, and a vacuum jacketed cryogenic liquid transfer
line with the appropriate flexible joints was installed to carry the
coolant to and from the detector inside the telescope.  Inside the
detector the coolant circulates through cooling loops attached
directly to the invar support bars to which the CCDs are attached,
insuring a temperature of all of the CCDs uniform to $\sim 1$\degr C.

The overall temperature can be regulated by controlling both the
temperature and the circulation rate of the coolant.  The camera's
overall heat load is $\sim 30$ watts, dominated by the heat radiated
in through the large vacuum window.  There are heaters attached to the
body of the camera to maintain the outside of the camera at ambient
temperature.

\subsection{The Vacuum System}
The camera dewar is operated at a vacuum of $\sim 10^{-4}$
torr, which is sufficient to keep the convective heat loss at a
negligible level.  The vessel is evacuated by a turbomolecular pump
preceeded by a roughing pump.  These pumps are not connected during
observations.  In fact, once the vessel is evacuated, the vacuum
usually lasts for weeks so the pumps are connected only periodically.

\subsection{The CCD Support Structure}

\begin{figure}
  \vspace*{2.7in}
  \includegraphics{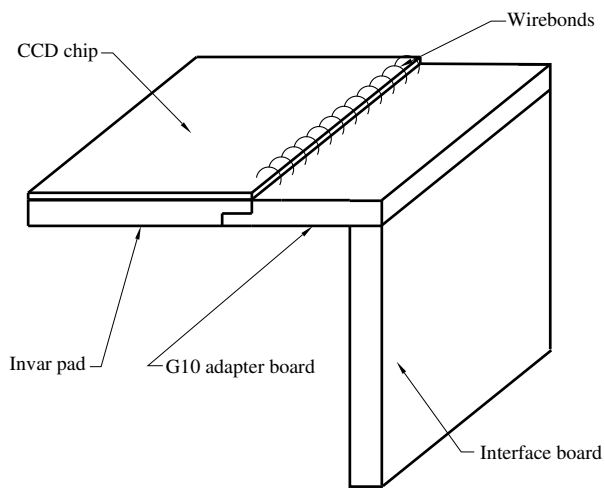}
  \caption{The CCD packages.}
  \label{f:CCD_package}
\end{figure}

The individual CCDs were packaged at Yale.  The CCDs were epoxied to a
$\sim 3$\mm\ thick invar plate approximately $3.1\cm \times 3.1\cm$
large.  A small circuit board was epoxied next to the invar plate and
the electrical connections from the CCD to this circuit board were
achieved by wire bonding to this board (see fig.~\ref{f:CCD_package}).
The CCD packages are mounted on four invar bars (fingers) each 0.25
inches thick, 1.0 inches wide and 11 inches long.  These four fingers
in turn are attached to a 14 inch diameter invar support plate 0.5
inches thick (see fig.~\ref{f:camera_front}).  This plate has a large
rectangular opening in the center to allow the electrical connections
from the CCDs to pass through to the preamplifier board located at the
back of the dewar under this plate.  The Schmidt telescope has a very
shallow depth of field ($\pm 25$\micr).  Therefore, all of the CCDs
have to be located in the focal plane to a precision smaller than
this.  All of the invar bars and the support plate were ground to
5\micr\ precision and all joints are spring loaded to ensure the
required precision.  The invar parts were coated with an
Armoloy\footnote{Armoloy is a thin, dense chromium coating that is
low friction, corrosion resistant and very hard.}
coating to prevent corrosion.
After the entire system was assembled, a detailed optical survey
showed that the front surfaces of all of the CCDs were in a plane with
an rms scatter of 11\micr.

The invar fingers pivot at one end and their rotational
position is controlled by cams located near the other end under the
invar support plate. The shafts of the four cams penetrate the back
plate of the dewar to the gears and stepping motors located outside of
the dewar.  The invar support plate is supported from the dewar back
plate by three standoffs which are 10\cm\ long hollow stainless steel
tubes to reduce the heat conduction to an acceptable level.

\subsection{Color Filters}

\begin{figure}
  \vspace*{3.5in}
  \includegraphics{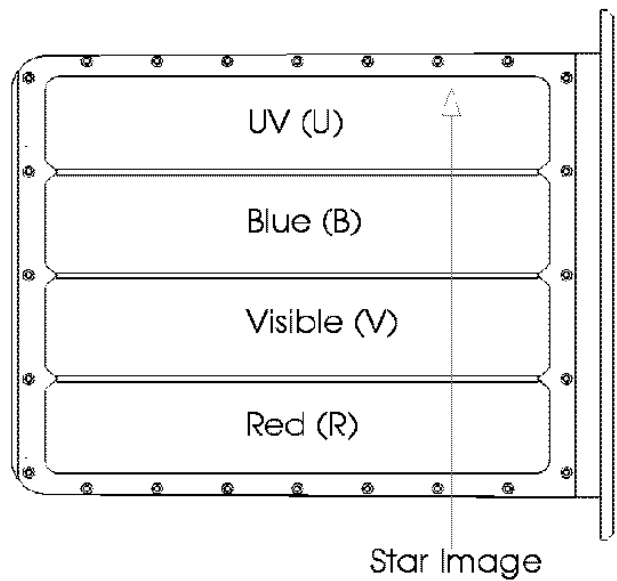}
  \caption{The filter tray with four typical color filters.}
  \label{f:filters}
\end{figure}

The filter box is located a few centimeters in front of the vacuum
window/field flattener lens.  Filter trays can be easily inserted or
removed manually. A filter tray consists of four filters of different
colors, each filter being 5.0\cm\ wide and 25.0\cm\ long.  Each
individual filter is located in front of one row of CCDs so that in
the course of a drift scan star images pass through each of the four
filters in turn so that data can be collected in four colors
essentially simultaneously.  A sketch of a filter tray is show in
figure~\ref{f:filters}.  Several filter trays exist and the individual
filters can be shuffled to make up filter trays in any desired
combination.  The available filter colors and their wavelength bands
are listed in table~\ref{t:filters}.  These filters were designed
specifically for the QUEST camera but they resemble the Johnson color
system \citep{Bessel1990} quite closely.  The optical thickness of the
different filters were designed such that the entire system is parfocal.

There is a grid of five nichrome wires located in the filter box with
the wires running along the edges of the individual filters. During
conditions of high atmospheric humidity a current is passed through
these heater wires to eliminate the condensation on the vacuum windows
and the filters.  This system has turned out to be quite effective.

\subsection{The Shutter Box}

The camera shutter is located in front of the filter box.  The
requirements on this shutter are to have a 12 inch diameter clear
opening but that no part of the shutter box extend beyond the 16 inch
diameter of the camera since the camera was located at the prime focus
of the telescope and obscuration of the light path was to be
minimized.  No such shutter was commercially available so a shutter
was custom designed and built for this purpose.  It is an iris type
shutter with 22 thin stainless steel shutter blades.  The shutter is
computer controlled and can open or close in about half a second.

\section{Detector Control System and Readout Electronics}
\label{sec:electronics}

\begin{figure}
  \vspace*{6.5in}
  \includegraphics{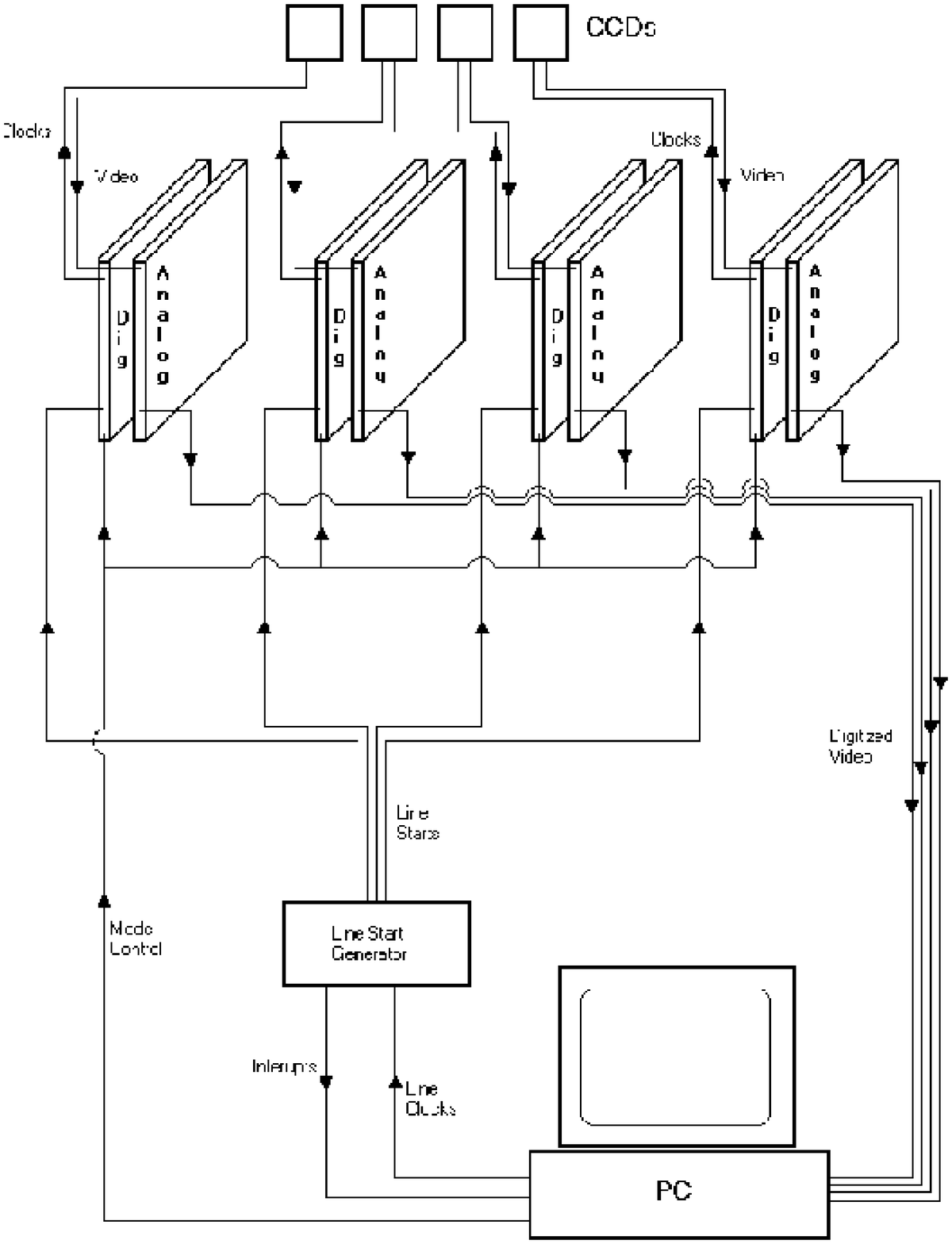}
  \caption{Schematic of the CCD detector read--out and control system.}
  \label{f:electronics}
\end{figure}

The electronics used to read out the camera are based on the CCD
controller developed for the U.~S.~Naval Observatory and the
Observatories of the Carnegie Institute of Washington by FHH Harris
Engineering.  A block diagram of the readout electronics is shown in
figure~\ref{f:electronics}.  The heart of the system is a two board
set, consisting of a digital board and an analog board, responsible
for the clocking and readout of a single CCD.  The phase clocks are
generated on the digital board as outputs of a state machine which is
implemented using field programmable logic devices (PLDs), allowing
considerable flexibility in the clock waveforms.  The analog board
amplifies and digitizes the CCD video output and provides the
digitized video to the data acquisition system, and is capable of
operation at a maximum readout rate of 100 kHz.  In addition to this
two board set for each CCD, the complete readout system includes a
line clock generator and an interface card to the data acquisition
computer.  As the system is implemented for this camera, the interface
card services four CCDs.  Control of the detector operating mode is
accomplished through a serial interface between the interface card and
the digital boards.  This mode control allows the selective readout of
an individual CCD, selection of drift scan or snapshot mode, and reset
and calibration of the control system.

The system interconnections are shown in figure~\ref{f:electronics}.
Each CCD video signal is provided to an analog board through a coax
cable.  All phase clocks and DC voltages required by the CCDs are
generated on the digital board, and provided to the camera via
shielded cable.  There are three groups of interconnections between the
controller electronics and the data acquisition system.  The digitized
video output is provided to the interface board via coax cable.  The mode
control is a slow serial interface between the interface card and the
digital boards, as described above.  Finally, each column of CCDs
receives a separate line start signal, which is generated by a
counter/timer card in the data acquisition computer.  Each of these
will be described in more detail below.

\subsection{Analog Board} \label{sec:ADC}

The analog board provides the amplification, processing, and
conversion to digital values of the output of the CCD detector.  The
analog board communicates with the associated digital board through a
backplane, located at the back of a rack in which the analog and
digital boards are mounted.  The input to the ADC section consists of
a double--correlated sample and hold, which is implemented as a
switched--input dual slope integrator.  The rather complicated sample
and hold circuit is required to eliminate sensitivity to variations in
the pre--charge levels on the CCD output node.  Prior to the delivery
of a charge packet on the CCD to the charge sense node, that node is
precharged to a positive potential.  After the precharging of the
sense node capacitance on the CCD via the reset clock applied from the
digital board, the value of the voltage potential of the sense node
has some uncertainty compared to previous precharge values.  This
uncertainty is due to the non--zero on--resistance of the reset switch
and the Johnson noise associated with that resistance.  The magnitude
of this voltage uncertainty can be of order 300 electrons, and must be
corrected for.  The ADC digitizes the output of the sample and hold to
16 bits.  The conversion rate of the ADC is 50 kHz.

\subsection{Digital Board}

The digital board provides the time-varying signals needed to operate
the CCD detector and its associated analog processing electronics on
the analog board.
As mentioned earlier, the phase clocks are
generated as outputs of a state machine having a total of 64 states,
with state transitions occurring at a rate of 4 Mhz.  The phase clocks
are output from clock drivers, which take as inputs the logic--level
outputs of the state machine, combining these with the DC voltage
outputs of a set of bias generators, and applying a simple RC low pass
filter, to produce the shaped, level--shifted signals required by the
CCD.

\subsection{Line Start Generator}

The camera is intended to be operated in drift scan mode.  In this
mode of operation, each column of CCDs, corresponding to CCDs at a
common declination, require a common line start clock, while different
columns will require slightly different line start rates, dictated by
the declination bands being viewed.  The method by which this is
implemented is to operate all columns at a common nominal line start
rate, but to drop line start cycles for each column at a rate which
gives the correct average line start rate for that column.  This
allows synchronous operation of the system, avoiding having large clock
transisitons ($\sim 10$ V) during the sensitive ($\sim 1$ mV) charge
readout time.

The line start generation circuitry consists of two components.  The
first is a commercial timer/counter board (NI PCI-TIO-10) running in the primary data
acquisition computer.  This board generates the correct nominal line
start rate, and, for each column, a cycle reject clock, which clocks
at the rate at which lines starts are skipped on that column.  Once
programmed, this board functions without intervention from the data
acquisition system.  The second module in the line start circuitry is
the line start generator.  This board accepts the raw clocks produced
by the timer/counter board, and generates the line starts used by the
controller boards.  These line starts are delayed by a fixed time
interval to allow the data acquisition computer time to prepare for
the line readout.  The line start generator also provides a set of prompt
interrupt lines to the data acquisition computer to notify the data
acquisition system of an incoming line.

\subsection{Data Acquisition Interface Card}

A 4--channel PC interface to the controller system has been developed
jointly at Princeton University and the U.~S.~Naval Observatory for
the Sloan Digital Sky Survey (SDSS) project.\footnote{We are grateful
to Jim Gunn (Princeton University) for supplying the PC interface
boards.}  This card provides a DMA interface to four CCD inputs, as
well as a serial output to the controller electronics which allows
control of the camera operational mode.

\subsection{Data Acquisition Computers}

\begin{figure}
  \vspace*{4.75in}
  \includegraphics{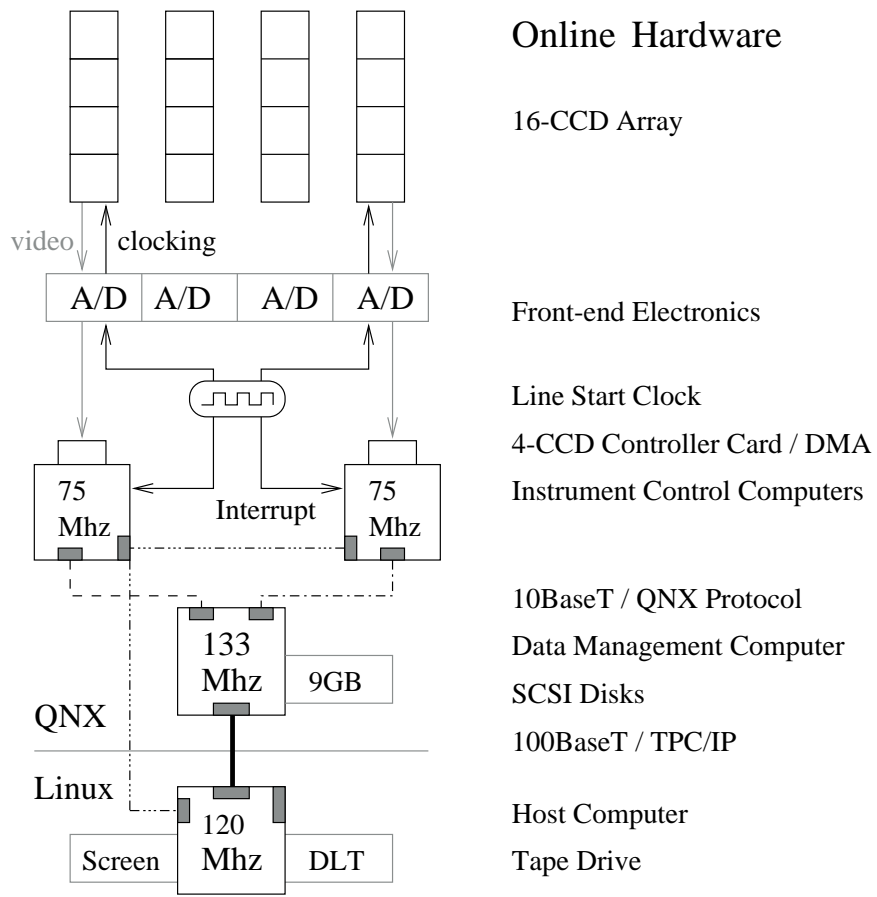}
  \caption{Block diagram of the data acquisition system.}
  \label{f:DAQ}
\end{figure}

The data acquisition system is built around standard IBM PC compatible
hardware.  For a full description see \citet{Sabbey1998}.  Each column
in the array (consisting of four CCDs), is controlled by one
Pentium--based CPU.  One controller board controls all 4 CCDs, sending
commands and receiving data. Because all chips in the column (indeed,
all in the detector) are clocked at the same rate, the entire data
acquisition can be run as one synchronous process.  This eliminates
many timing headaches.  The data is stored during the night on disks,
and written to DLT tapes at the end of a night's observing.  The four
data acquisition computers run on the QNX operating system and are
linked through two data storage computers to one central controlling
computer via Ethernet.  The control computer, running Linux as its
operating system, organizes the data taking process, and monitors the
quality of data during the night, using samples sent from the data
taking machines over the ethernet connection.  A block diagram of the
data acquisition hardware is shown in Figure~\ref{f:DAQ}.

\section{Data Rates and Data Analysis Software}
\label{sec:software}
\subsection{Data Rates}

As discussed in \S\ref{sec:ADC} above, the analog to digital
converter (ADC) outputs 16 bits for the signal from each pixel.  The
calibration of the electronics varies slightly from channel to channel
but is in the vicinity of 1 electron per ADU (analog to digital
conversion unit).  In the drift scan mode the entire $64 \times 10^{6}$
pixels of the array are read out once every 140 seconds with 2 bytes
per pixel, resulting in a data rate of approximately 1 megabyte per
second.  This data is stored on disks and written on a DLT tape
in the morning after the observation night.  A clear 8
hour night produces about 28 gigabytes of data.  With appropriate
compression \citep{Sabbey1999b} this will fit on a single DLT IIIXT tape.
These tapes are the raw data input for the offline processing.

The software to process such a large volume of data is not trivial.
The QUEST collaboration has developed three different software
pipelines to analyze this data: a) the photometric pipeline to analyze
data taken with the multicolor filters without the objective prism; b)
the spectroscopy pipeline to analyze data taken with the objective
prism; and, c) the supernova pipeline, a highly specialized program
used in the search for type Ia Supernovae and other variable objects.
The output of these pipelines are typically object catalogs which form
the starting points of more specialized data analysis programs.  
Examples of these programs include searching for variable objects such
as RR~Lyrae stars, TNOs, quasars, etc., or identifying quasars by their colors
(in direct data) or broad emission lines (in objective prism data).

\subsection{The Photometry Pipeline}
This program was developed to analyze data taken in direct mode
(without the objective prism) using multicolor filters --- typically
U, B, U, V, or R, B, R, V.  Data from repeated scans of the same area
of the sky are co--added.  One color, usually V or R, is selected as
the lead color and is used to find objects.  The locations of the
objects found in the lead color are translated to the coordinate
system of each of the other colors for each night of data and
photometry, using both aperture photometry and PSF (Point Spread
Function) fitting, is carried out to obtain magnitudes in each color.
Landoldt standards \citep{Landolt92} or secondary standard stars are used for
photometric calibration.  The astrometric calibration is carried out
using the USNO A2.0 catalog \citep{Monet1999}; the typical resulting precision
is about $0.2\arcsec$.  A more detailed description of this pipeline will
be published in a separate article.

\subsection{The Spectroscopy Pipeline}
\begin{figure}
  \vspace*{4in}
  \includegraphics{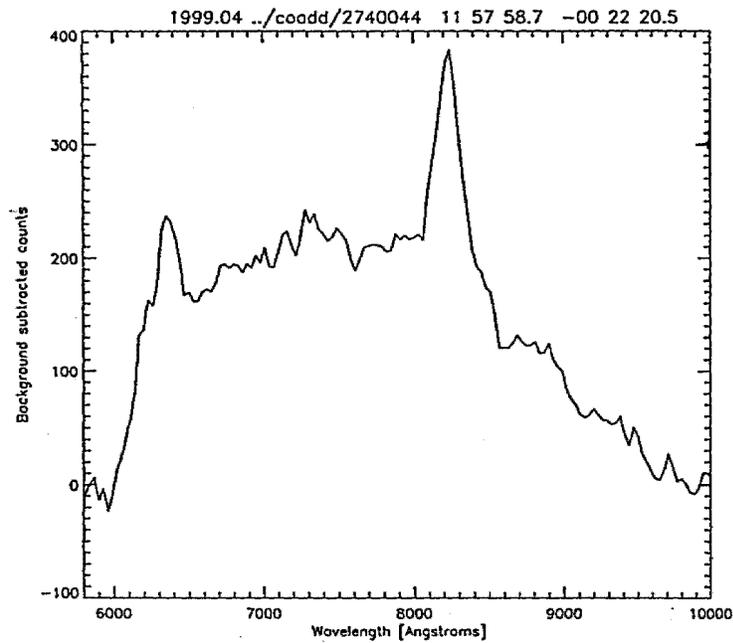}
  \caption{The objective prism spectrum of a $z=0.25$ QSO with two broad
emission lines identified by the spectroscopy analysis program.}
  \label{f:sabbey_QSO}
\end{figure}

This program has been developed \citep{Sabbey1999a} to analyze drift
scan data taken with the objective prism.  Data from repeated scans of
the same region of the sky are co--added.  From the co--added data,
spectra of individual objects are extracted, background subtracted and
calibrated. The program then automatically examines each spectra and
selects objects with prominent emission lines from which redshifts can
be determined \citep{Sabbey_thesis}.  A typical objective prism
spectrum with two broad emission lines, identified by the program is
shown in figure~\ref{f:sabbey_QSO}.

\subsection{The Supernova Pipeline}
This program was developed by the Supernova Cosmology Project
\citep[see, for example,][]{Perlmutter1999} and was adapted to work
with the output of the QUEST Camera in drift scan mode. In this
program, data from ``discovery'' nights and from ``reference'' nights
(typically about two weeks earlier than the discovery nights) are
convolved to the same seeing and normalized to the same intensity
scale.  The reference nights are then subtracted pixel--by--pixel from
the discovery nights.  The vast majority of the objects disappear in
the subtraction.  Objects with significant residuals in the
subtraction are examined visually as candidates for variable objects
such as supernovae, trans--Neptunian objects, asteroids, etc.

\section{Performance of the Apparatus}
\label{sec:performance}

The QUEST camera was commissioned in the Venezuelan Schmidt telescope
in 1997.  We have had three observing seasons from November 1998 to
May of 2001.  Using these observations, the performance of the
combination of the camera, the telescope, and the site of Llano del
Hato have been fairly well characterized.

\subsection{Alignment Procedure}
During the initial commissioning phase, the plane of the front
surfaces of the CCDs had to be aligned to be parallel to the focal
plane to within the $\pm 25$\micr\ depth of field of the telescope.
This was done by tilting adjustments around the two relevant
rotational degrees of freedom.  The method used was to precisely
measure the optimal focus position of each individual CCD.  The most
sensitive method used a Hartmann mask (a sheet at the entrance of the
telescope with a pattern of small holes) which produces double images
when the CCDs are not in focus.  The image separation is proportional
to the offset from the best focal position. By taking two exposures,
one well below and the second well above focus, the precise focal
position of each CCD could be determined.  Initially, when the CCD
plane was not exactly parallel to the telescope focal plane, a
systematic tilt in these focal positions was observed.  After several
interations of adjusting the tilt angle and repeating the Hartmann mask
test, the CCD plane was brought into the telescope focal plane well
within the depth of field.  A check of the adequacy of this alignment
is to examine the full width half maximum (FWHM) of the Point Spread
Function (PSF) of the images in the individual CCDs at a single
focal position of the entire camera.  Such a set of measurements is
shown in figure~\ref{f:FWHM}.  No systematic variation of the FWHM across the
diagonal of the full image plane is observable.  This alignment
procedure, carried out during commissioning, did not have to be
repeated except once after the camera was taken out of the telescope
for maintenance.

The rotational position of each of the four fingers and the readout
clocking rate synchronization has to be optimized or checked every
time the declination of the drift scan is changed.  The expected
finger position and clocking rates are calculated and set by the
control computer for the desired declination. At the beginning of
observations of a new value of the declination, these settings are
optimized by varying the finger positions and the clocking rates in
small steps around the expected values.  If the finger position is
off, the images are elongated in the North--South direction, and if
the clocking rate is off, the images are elongated in the East--West
direction. The optimum settings are those that produced the smallest,
round images.  After some experience with the camera, we learned to
trust the computer settings and this optimization procedure was
undertaken only when the images did not have the expected point--like
shape.

\subsection{Seeing Quality}

\begin{figure}
  \vspace*{3.4in}
  \includegraphics{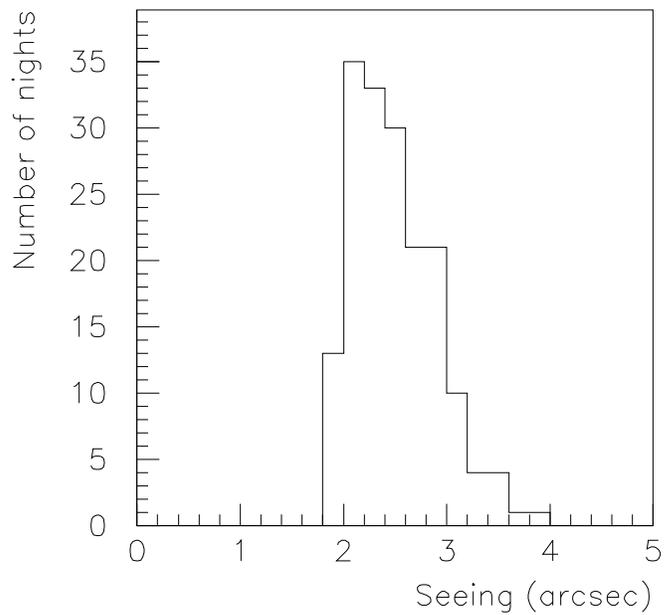}
  \caption{Distribution of the image FWHM of data taken in March 1999 and 2000.}
  \label{f:seeing}
\end{figure}

During commissioning, the seeing FWHM was in the vicinity of 3\arcsec.  A
considerable effort was made to improve the ventilation of
the dome, putting thermal enclosures vented to the outside around the
electronics box and the refrigeratation unit located on the floor near
the telescope, and putting servo controlled heaters on the camera body
inside the telescope. After these improvements, the seeing was closer
to 2\arcsec.  In the drift scan mode with effective 140 second
exposure times, the best seeing observed was 1.8\arcsec.  The
distribution of the FWHM during the month of March (1999 and 2000
combined) is shown in figure~\ref{f:seeing}.  The median seeing was
around 2.4\arcsec.  The best seeing in the non-drift scan point
and stare mode, with 5 second exposures, is shown in
figure~\ref{f:FWHM}.  The best seeing for these short
exposures was 1.5\arcsec.  The degradation from $\sim 1.5\arcsec$ to
$\sim 1.8\arcsec$ is consistent with what we expect from the
effects of drift scanning discussed in \S~\ref{sec:operation} above.

\subsection{Readout noise and sky background}
The rms fluctuations (noise) on the output amplitudes of the CCDs have
contributions from the read noise inherent in the CCDs, the dark
current in the CCDs, the noise introduced by the electronics, pick up
noise in the cabling and cross talk, fluctuations in the sky
background, and finally the statistical fluctuations of the star light
signal itself.  The combination of the read noise and dark current in
the CCDs, the electronics noise and the pickup noise, was measured by
looking at the rms spread of the signals of dark exposures.  This
combined noise varied from 9 to 21 electrons with a median of 13
electrons per pixel.

The sky background depends strongly on the color filters used.  For a dark
night, it is smallest, around 20 electrons/pixel/140 seconds with the U
filter, and largest, around 600 electrons/pixel/140 seconds with the R filter.
Typical values of the sky background with the different filters used
are given in table~\ref{t:sky_background}.  These backgrounds, of
course, also vary a fair amount with the phase of the moon and the
atmospheric conditions.  Under normal conditions we are read noise
limited in the U and B filters, but sky noise limited in the V and R
filters.

\subsection{Limiting Magnitudes and Photometric Errors}

\begin{figure}
  \vspace*{5in}
  \includegraphics{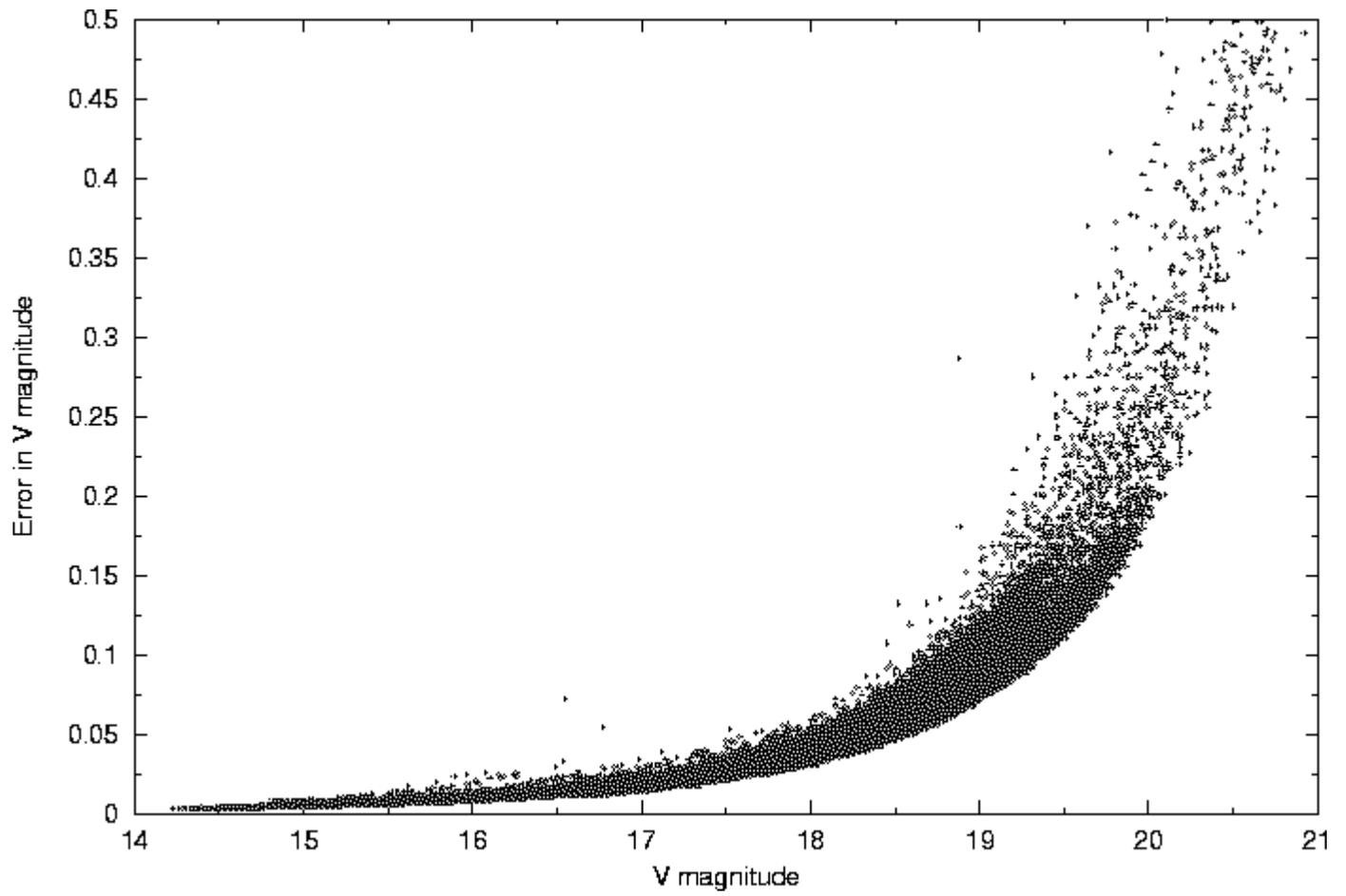}
  \caption{Plot of the error on the magnitude versus the magnitude with
the V color filter.}
  \label{f:mag_err}
\end{figure}

We take the limiting magnitudes to be the magnitude of objects for
which the signal to noise is larger than 10 to 1, or the total error
on the magnitude of the object is less than 0.1 magnitudes.  The
limiting magnitudes depend strongly on the color filter used, and also
on the phase of the moon and the atmospheric conditions.  We estimate
the limiting magnitude by plotting the error in the measured magnitude
versus the magnitude with a given filter.  An example of such a plot
is shown in figure~\ref{f:mag_err}.  This plot is for a single CCD
with 140 second drift scan exposures over a number of nights with dark
moon, so the spread in the points is primarily due to atmospheric
conditions.  The curve crosses the 0.1 mag error line between
magnitudes 18.8 and 19.4 which we take to be the limiting magnitude.
The typical limiting magnitudes are given in
table~\ref{t:sky_background}.  At the bright end, stars saturate the
CCDs around magnitude 13.

The assigned photometric errors on the measured magnitudes were
calculated in the photometric pipeline program.  The correctness of
these assigned errors was checked by looking at repeated measurements
of the same objects observed in about 20 repeated scans of the same
area of the sky in March of 1999 and 2000.  The photometric program
corrected for variations in atmospheric extinction from night to
night.  After these corrections, the rms spread of the actual
individual measurement of the same object agreed well with the error
calculated by the program.  For bright objects, around magnitude 15 or
16, the errors were as low as 0.002 magnitudes.  The errors discussed
above do not include the photometric calibration errors; these depend on
the number and proximity of the standard stars used in the calibration.

\subsection{Scientific Results}

During the first three years of operation of the CCD camera we
obtained a survey of 700 square degrees with the objective prism, a
survey of 1000 square degrees with color filters and a 250 square
degree variability survey with repeat time scales varying from twice a
night to three years.  Some of the scientific results obtained include
a quasar correlation and large scale structure study
\citep{Sabbey2001}, discovery of the optical counterpart of a gamma
ray burster \citep{Schaefer99}, discovery of a new minor planet
2000~EB173 \citep{Ferrin2000}, an RR~Lyrae survey \citep{Vivas2001},
a star formation and T~Tauri star study \citep{Briceno2001}, a sample of
about 30 type Ia supernovae \citep{paper_6}, and a sample of about
5000 quasars indentified by a variety of techniques \citep{paper_7}.

\begin{figure}
  \vspace*{6in}
  \includegraphics{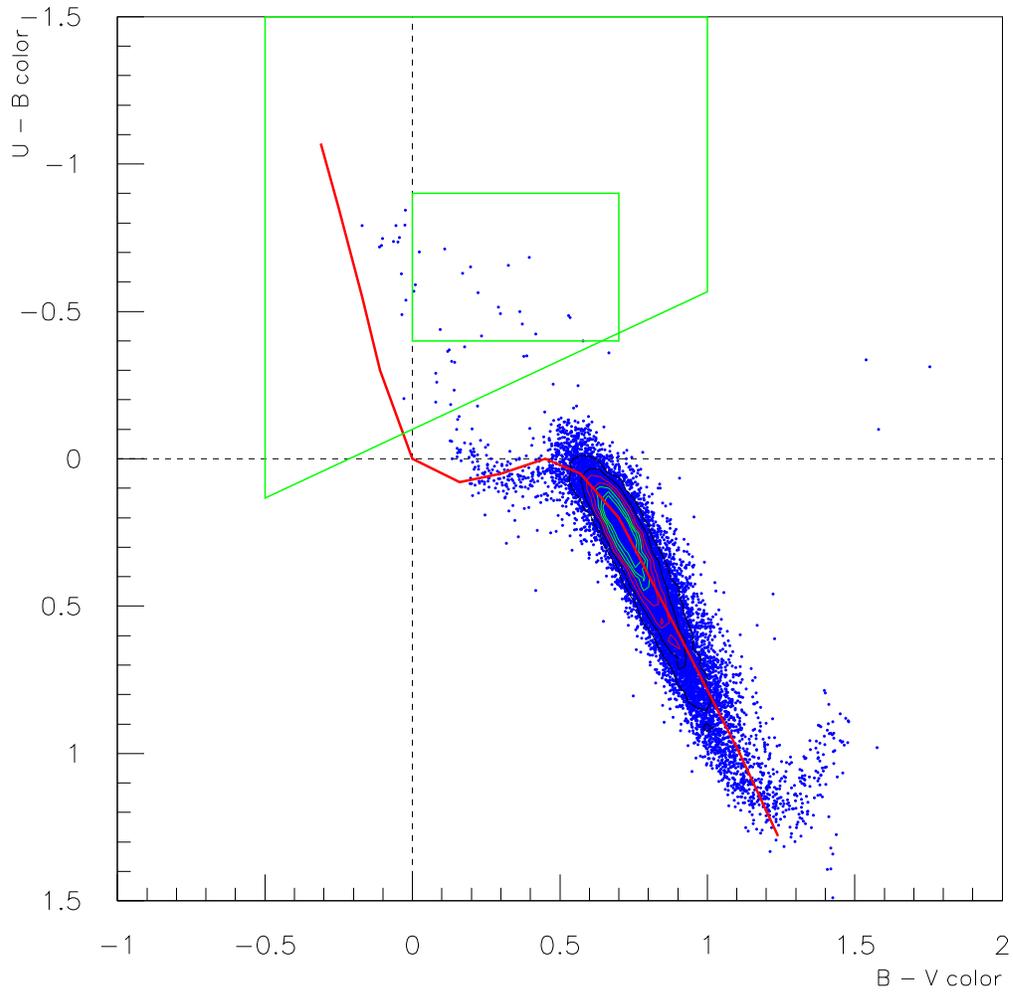}
  \caption{The UBV color--color diagram, showing clustering along the
main sequence line and quasar candidates in the expected region.}
  \label{f:color-color}
\end{figure}

As an example of the quality of the data, figure~\ref{f:color-color}
shows a UBV color--color diagram of a typical observation.  There is a narrow
concentration where we expect the main sequence stars, the overall
random background is small, and there is a small concentration in the
region where we expect quasars with redshifts below 2.2.
Spectroscopic followup has demonstrated that the efficiency of this
quasar sample is around 65\% and comparison with catalogs of known
quasars indicate a completeness around 70\%.

\clearpage
\renewcommand\thepage{}
\bibliographystyle{apj}        

\clearpage

\clearpage

\begin{table}
\begin{center}

\begin{tabular}{|l|l|} \hline
Parameter		& Value		\\ \hline \hline
Clear aperture diameter	& 1.00\m	\\
Mirror diameter	        & 1.52\m	\\
Focal length	        & 3.03\m 	\\
$f$ ratio	        & $f/3$  	\\ 
Plate scale	        & 15\micr/arcsec (67 arcsec/mm)	\\
Corrector plate glass	& UBK7     	\\
Objective prism:        &          	\\
\hspace*{.25in}Aperture Diameter	& 1.0\m   	\\
\hspace*{.25in}Wedge angle	& 3.3\degr 	\\
\hspace*{.25in}Dispersion	& 650\angst/\mm\ at 4350\angst \\
\hspace*{.25in}Glass		& UBK7  \\
Latitude of observatory	& 8\degr\ 47\arcmin\ North \\
Longitude of observatory & -70\degr\ 52\arcmin\ 0\arcsec \\
Elevation	        & 3600\m\    \\ \hline  
\end{tabular}
\end{center}
\caption{\label{t:properties}Properties of the Venezuelan Schmidt Telescope}
\end{table}

\begin{table}
\begin{center}
\begin{tabular}{|l|l|} \hline
Parameter		& Value \\ \hline \hline
Number of CCDs		& 16		\\
Array size, CCDs	& $4 \times 4$	\\
For each CCD:	&			\\
\hspace*{.25in}pixel size	& $15\micr \times 15\micr$\\
\hspace*{.25in}Number of pixels	& $2048 \times 2048$	\\
\hspace*{.25in}Pixel size on sky	& $1\arcsec \times 1\arcsec$\\
Array size, pixels	& $8192 \times 8192$ pixels	\\
Array size, cm		& $12.6\cm \times 18.2\cm$	\\
Array size, on sky	& $2.3\degr \times 3.5\degr$	\\
Effective area		& 5.4 square degrees		\\ \hline
\end{tabular}
\end{center}
\caption{\label{t:camera_prop}Properties of the QUEST camera}
\end{table}

\begin{table}
\begin{center}
\begin{tabular}{|l|l|}  \hline
Parameter		& Value		\\ \hline \hline
CCD size, mm		& 31.28mm $\times$ 31.75 mm	\\
CCD size, pixels	& 2048 $\times$ 2048		\\
Pixel size		& $15\micr \times 15\micr$\\
CCD type		& Front illuminated\\
Clocking rate in drift scan mode:	&\\
\hspace*{.25in}parallel clock	& $\sim$ 15 rows/sec\\
\hspace*{.25in}serial clock	& 50 kilohertz\\
Read noise		& $\sim$ 10 electrons\\
Dark current at 20\degr C	& 250 pA/cm$^{2}$\\
Full well capacity	& 30,000 to 60,000 electrons\\
Quantum efficiency:	&			\\
\hspace*{.25in}at 7000\angst	& $\sim$ 45\%	\\
\hspace*{.25in}below 4000\angst	& 5 to 10\%	\\  \hline
\end{tabular}
\end{center}
\caption{\label{t:CCD_prop}Characteristics of the CCDs used in the
camera}
\end{table}

\begin{table}
  \begin{center}
  \begin{tabular}{|l|l|} \hline
Color	& Wavelength Range (\angst)\\ \hline \hline
U	& 3300 --- 4000		\\
B	& 3900 --- 4900		\\
V	& 5050 --- 5950		\\
R	& 5900 --- 8100		\\
I	& 7800 --- 10,200	\\
H$\alpha$ & 6520 --- 6600	\\
Broad B	& 4000 --- 6500		\\
Broad R	& 6500 --- 9000		\\ \hline
\end{tabular}
\end{center}
  \caption{Color filters available for the QUEST camera.}
  \label{t:filters}
\end{table}

\begin{table}
\begin{center}
\begin{tabular}{|c|c|c|} \hline
	& Typical Sky Background	& Limiting Magnitude\\ 
Filter	& Electrons/pixel/140 seconds	& S/N $\geq$ 10\\ \hline
U	& 20	& 16.5\\
B	& 200	& 18.5\\
V	& 300	& 19.2\\
R	& 600	& 19.5\\ \hline
\end{tabular}
\end{center}
\caption{Typical sky backgrounds for a dark night and limiting magnitudes
for a single 140 second exposure.}
\label{t:sky_background}
\end{table}

\end{document}